\documentclass{article}
\usepackage{PSIP2003,amsmath,epsfig}

\def\adj{^{\dagger}}

\title{Component separation for Cosmic Microwave Background data: A blind approach based on spectral diversity}
\name{Guillaume Patanchon$^1$, Hichem Snoussi$^2$, Jean-Fran{\c c}ois Cardoso$^3$, Jacques Delabrouille$^1$}
\address{$^1$ PCC --- Coll{\`e}ge de France, 11, place Marcelin Berthelot, F-75231 Paris, France\\ 
$^2$ L2S, Sup\'elec, Plateau de Moulon, 91192 Gif-sur-Yvette Cedex, France\\
$^3$ CNRS/ENST --- 46, rue Barrault, 75634 Paris, France
}
\begin{document}
%
\maketitle
\begin{abstract}
We present a blind multi-detector multi-component spectral matching method for
all sky observations of the cosmic microwave background, working on the
spherical harmonics basis. The method allows to estimate on a set of
observation maps the power spectra of various components present in CMB data, 
their contribution levels in each detector and the noise levels. The method accounts for the
instrumental effect of beam convolution. We have implemented the method on all 
sky Planck simulations containing five components and white noise, including beam smoothing effects.
\end{abstract}


\section{Introduction}
\label{sec:intro}

The precise measurement of Cosmic Microwave Background (CMB) anisotropies is
one of the main objectives of modern cosmology. The small temperature
fluctuations of the CMB with respect to the pointing direction on the sky
reflect the primordial density perturbations in the young Universe. The spatial
power spectrum of those fluctuations depend on a set of important parameters, known as cosmological parameters. The accurate estimation of CMB power spectrum
is thus of prime importance in cosmology. The Planck satellite (to be launched in 2007) will map
whole sky emission with unprecedented signal to noise ratio.
\\ The accuracy of CMB power spectrum measurements are limited by the other
astrophysical emissions present in the sub-millimeter range of the spectrum.
Those emissions, called foregrounds (as they are emitted in front of the CMB),
are of different origin. Some of them originate from within our own Galaxy, as 
the dust and synchrotron emissions, others are of extra-galactic
origin as the Sunyaev-Zel'dovich effects. All depend on the wavelength.\\

It is thus important for a precise measurement of the CMB power spectrum
to deal with the various foregrounds present with the CMB anisotropies. The
availability of several detectors operating in several bands (10 for Planck
ranging from 30 to 850 GHz) allows to distinguish the various contributions. Component separation methods has been
adressed by a number of authors \cite{Wiener,wiener96tegmark,MEM,allskyMEM,allskywienerPrunet,MEMWave,FICA,FICAallsky}. The standard approach consists in
producing the cleanest maps of the various components, followed by estimation
of the power spectrum of the
CMB from the separated CMB map. This approach is not fully satisfactory for two
reasons. First, component separation requires prior knowledge of the
electromagnetic spectra of the components which are not all very well known.
Second, it would be preferable to
estimate the CMB power spectrum in one step by jointly analysing the
observation maps.\\ A new approach to process multi-detector multi-component
data has been developed in papers \cite{paper1} and \cite{jfc}. The method is based on
likelihood maximization in the Whittle approximation and takes additive
noise into account (in the CMB observations, a significant amount of noise is
expected). The spectral diversity of the various components is used.
The method has been implemented on the domain of small sky maps, in the flat
sky approximation.\\ In this paper, we have refined the multi-detector
multi\--compo\-nent (MDMC) spectral matching method in several aspects to account for
distinctive CMB observation features. First, as current observations cover as
large a portion of the sky as possible, we have adapted the method for a spherical
harmonic expansion of all sky maps in order to exploit all the
information contained in the maps. Second, we account for the finite spatial
resolution of the detectors, to take into account the fact that the sky is
not seen with the same resolution by all detectors. Finally, our method allows
for the inclusion of some physical knowledge about the components. Fixing some
parameters, the likelihood is maximized over the other parameters. This
operation allows, for example, to break degeneracies due to components having
similar spatial power spectra.\\ We have implemented the method on full sky
simulated Planck observations. We compare the estimated CMB power spectra in the
blind approach and in the ideal case where one has a perfect knowledge of the
component emission laws.

\section{Model of sky emission}
\label{sec:model}

The key assumption is that the sky emission at a given frequency is the linear
superposition of astrophysical components. In addition, we assume that the emission
laws of the various components do not depend on the position on the sky. 
The signal measured by a detector is the sky emission convolved by a beam shape
(depending on the detector), plus an additive noise. Assuming a symmetric beam
shape, the observation by a detector $d$ is given by~:
\begin{equation}
  x_d(\vec{r})=\sum_{c=1}^{N_c}~A_{dc}~.~\int b_d(|\vec{r}-\vec{r'}|)~.~s_c(\vec{r'})~d\vec{r'}~+~n_d(\vec{r})
\label{eq:gen_form_obs}
\end{equation}
$\vec{r}$ is the direction of observation on the sky, 
$s_{c}$ is the emission template for source $c$, $n_d$ represents the 
noise of the detector $d$, $b_d(|\vec{r}-\vec{r'}|)$ represents the beam, and
depends only on $|\vec{r}-\vec{r'}|$ (or on $\vec{r}.\vec{r'}$) for symmetric
beams and $A$ is the mixing matrix. Each element of the mixing matrix results
from the integration of the emission law of one component over one detector
frequency band.

 A natural basis for the application of the MDMC
spectral matching method, exploiting the components spectral
diversity is the spherical harmonics basis. We write the decomposition of the
signal in this basis~:
\begin{equation}
  x_d(\vec{r}) = \sum_{l=0}^{\infty}\sum_{m=-l}^l ~{x_d}(l,m)~Y_l^m(\vec{r})
\end{equation}
where $Y_l^m(\vec{r})$ are the spherical harmonic functions. 
 Coefficients ${x_d}(l,m)$ can be easily computed using the orthogonality of spherical harmonic functions~:
\begin{equation}
  {x_d}(l,m) = \int_{4\pi} x_d(\vec{r})Y_l^m(\vec{r})^* d\Omega
\label{eq:decomp}
\end{equation}

The convolution between a symmetric beam and the
signal in real space becomes a product in the spherical harmonic basis. Then,
we obtain the observation coefficients combining equations \ref{eq:gen_form_obs} and \ref{eq:decomp}
\begin{equation}
 {x_d}(l,m) = b_d(l) \sum_{c=1}^{N_c} A_{dc}~s_c(l,m) + n_d(l,m)
\label{eq:obsmodel}
\end{equation}
where $b_d(l)$ is the Legendre Polynomial expansion of \\
${\rm b_d}({\rm cos}\Theta)$, $\Theta$ being the angular distance from the
center of the beam so that ${\rm cos}\Theta=\vec{r}.\vec{r'}$.
For a Gaussian beam, $b(l) \simeq \exp(-\sigma_b^2~l(l+1)/2)$
and $\sigma_b = \Theta_{\rm beam}/\sqrt{8\rm{ln}2}$, where $\Theta_{\rm beam}$ is the full width at half maximum of the beam.\\ Let us consider the diagonal
matrix $B(l)$ such that the diagonal element $B_{dd}(l) = b_d(l)$. We define the new coefficients
${x'}(l,m)=B(l)^{-1} x(l,m)$. It is useful to write these ``deconvolved'' 
observations coefficients in a matrix form~:
\begin{equation}
  {x'}(l,m) = A s(l,m) + B(l)^{-1}n(l,m)
\label{eq:ourmodel}
\end{equation}
The introduction of $x'$ will be justified later.

{\bf Spectral statistics}

We now need to compute the spectral statistics of the observations ${x'}(l,m)$,
given by $R_{x'}(l,m) = \langle x'(l,m)  x'(l,m){^\dagger} \rangle$, where
$\cdot\adj$ denotes transpose-conjugation.
\begin{equation}
  R_{x'}(l,m) = A C(l,m) A\adj + B(l)^{-1}N(l,m)B(l)^{-\dagger}
\label{eq:covmat}
\end{equation}
$C(l,m)$ and $N(l,m)$ are respectively the component and the noise
covariance matrices. Statistical independence between components
implies that $C_{lm}$ is a diagonal matrix. We also assume that the noise is
white and independent across detectors, so that~: $N(l,m) = diag(\sigma_1^2,...,\sigma_{d}^2)$.
The main reason to work with variable $x'$ is that we can average the
covariance matrices $R_{x'}(l,m)$ over bins and preserve the simple structure of equation
(\ref{eq:covmat}) for the signal part. Let us define the following particular averaging over bins:
\begin{equation}
  R_{x'}(q) = {\frac 1 {n_q}} \sum_{l=l_{\rm min}(q)}^{l_{\rm max}(q)} \sum_{m=-l}^l R_{x'}(l,m)
\end{equation}
Here $q=1,Q$ is the spectral bin index, the modes $(l,m)$ contributing to $q$ are
such that $l_{\rm min}(q)<l<l_{\rm max}(q)$.
Since $m$ can vary between $-l$ and $l$, the number of modes in each bin $q$
is $n_q = \sum_{l=l_{\rm min}(q)}^{l_{\rm max}(q)}(2l+1)$. The reason for the choice of such averaging
is that the spectral covariance matrices of isotropic components on the sky
(we have strong reasons to think that cosmological components are isotropic)
do not depend on the parameter $m$, so $C(l,m)=C(l)$. The average covariance matrices are~:
\begin{equation}
  R_{x'}(q) = A C(q) A\adj + M(q)
\label{eq:sptheo}
\end{equation}
where $M(q) = 1/n_q \sum_{l=l_{\rm min}(q)}^{l_{\rm max}(q)}(2l+1) B(l)^{-2}N$ is a diagonal matrix.

The band-averaged spectral covariance matrices are estimated by~:
\begin{equation}
  \widetilde R_{x'}(q) = {\frac 1 {n_q}} \sum_{l=l_{\rm min}(q)}^{l_{\rm max}(q)} \sum_{m=-l}^l {x'}(l,m){x'}(l,m)\adj
\label{eq:spexp}
\end{equation}
which is real valued since ${x'}(l,-m) = {x'}(l,m)\adj$ for real data.

\section{The MDMC method}
\label{sec:Method}

The aim of the MDMC spectral matching method is to obtain an estimate of
different parameters of the model, of relevance in astrophysics and cosmology, without the need
of prior information. Those parameters are
the mixing matrix $A$, the band averaged component power spectra $C_q$ and
the noise covariance $N$. They are collectively referred to as \\$\theta=\{A,C(q),N\}$.\\
The method is based on minimazing the mismatch between the empirical spectral covariance
matrices of the observations (equation \ref{eq:spexp}) and their expected values which depends on the
mixing parameters (equation \ref{eq:sptheo}). The mismatch is quantified by the average divergence measure between two matrices~:
\begin{equation}
   \Phi(\theta) = \sum_{q=1}^Q n_q D(\widetilde R_{x'}(q),R_{x'}(q))
\label{equation:phi}
\end{equation}
where $D(R_1,R_2)$ is a mesure of the divergence (the Kullback divergence) between two positive matrices and $n_q$
is the number of modes in each band $q$. Assuming that the spherical harmonics
coefficients of the components are random realizations of Gaussian field of
variance $R_{x'}(q)$ and are uncorrelated, the log-likelihood (up to irrelevant
factor) take the same form as in equation \ref{equation:phi} (in the frame of the Whittle approximation). The divergence is given
in this case by~:
\begin{equation}
   D(R_1,R_2) = {\rm tr} (R_1R_2^{-1}) - {\rm log~det} (R_1R_2^{-1})-m_d
\end{equation}
The parameter estimate is given by~$\hat\theta = {\rm argmin}_{/\theta}\Phi(\theta)$.
The connection with the log-likelihood
guarantees good statistical properties of the estimates (at least asymptotically).

{\bf Optimization algorithm:}

The optimization is made using an Expectation-Maximi\-zation
(EM) algorithm, and completed by quasi-Newton algorithm (BFGS) in order to
accelerate the convergence. The formal algorithm described in \cite{paper1} is
used, with minor changes (the equations change slighly when we account for
the beam smoothing effect). 



{\bf Degeneracies}

As seen from equation \ref{eq:sptheo}, one can exchange a factor between the mixing matrix $A$ and the component
power spectra $C(q)$ without changing the result of the equation. We fix this degeneracy in the EM step by fixing the norm of each
column of $A$ to unity; the power spectra are adjusted accordingly. In the 
quasi-Newton step, a penalty term is added to $\Phi$ which penalizes the
deviation of each column of $A$ with a minimum penalization at unity.

\section{Application}
\label{sec:application}

We now turn to the application of the MDMC spectral matching method on synthetic data.

\subsection{Simulated Planck data}

\begin{figure*}[p]
\begin{minipage}{2\columnwidth}
   \begin{center}
\includegraphics[draft=false,width=4.7cm,angle=90]{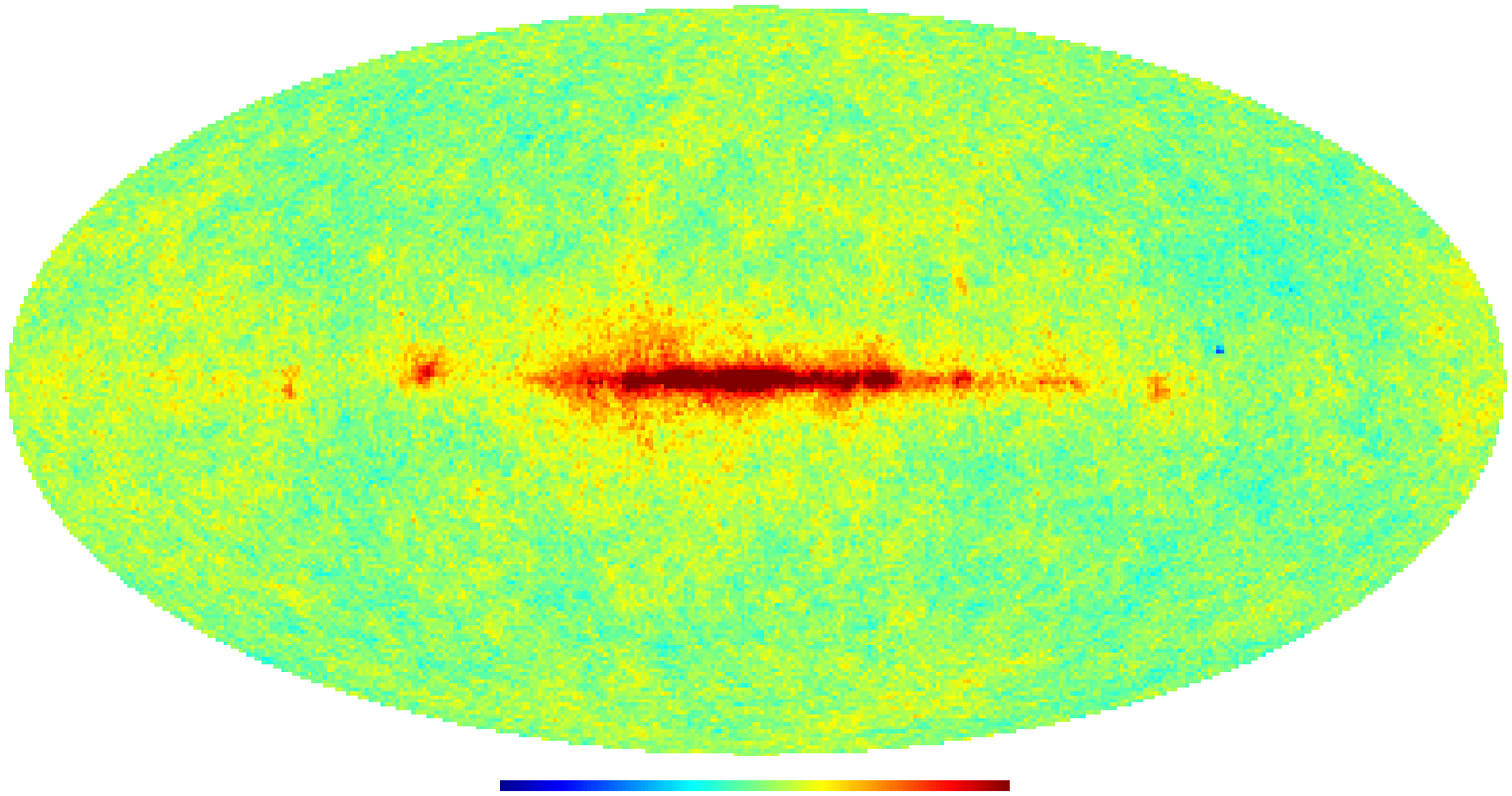}
\includegraphics[draft=false,width=4.7cm,angle=90]{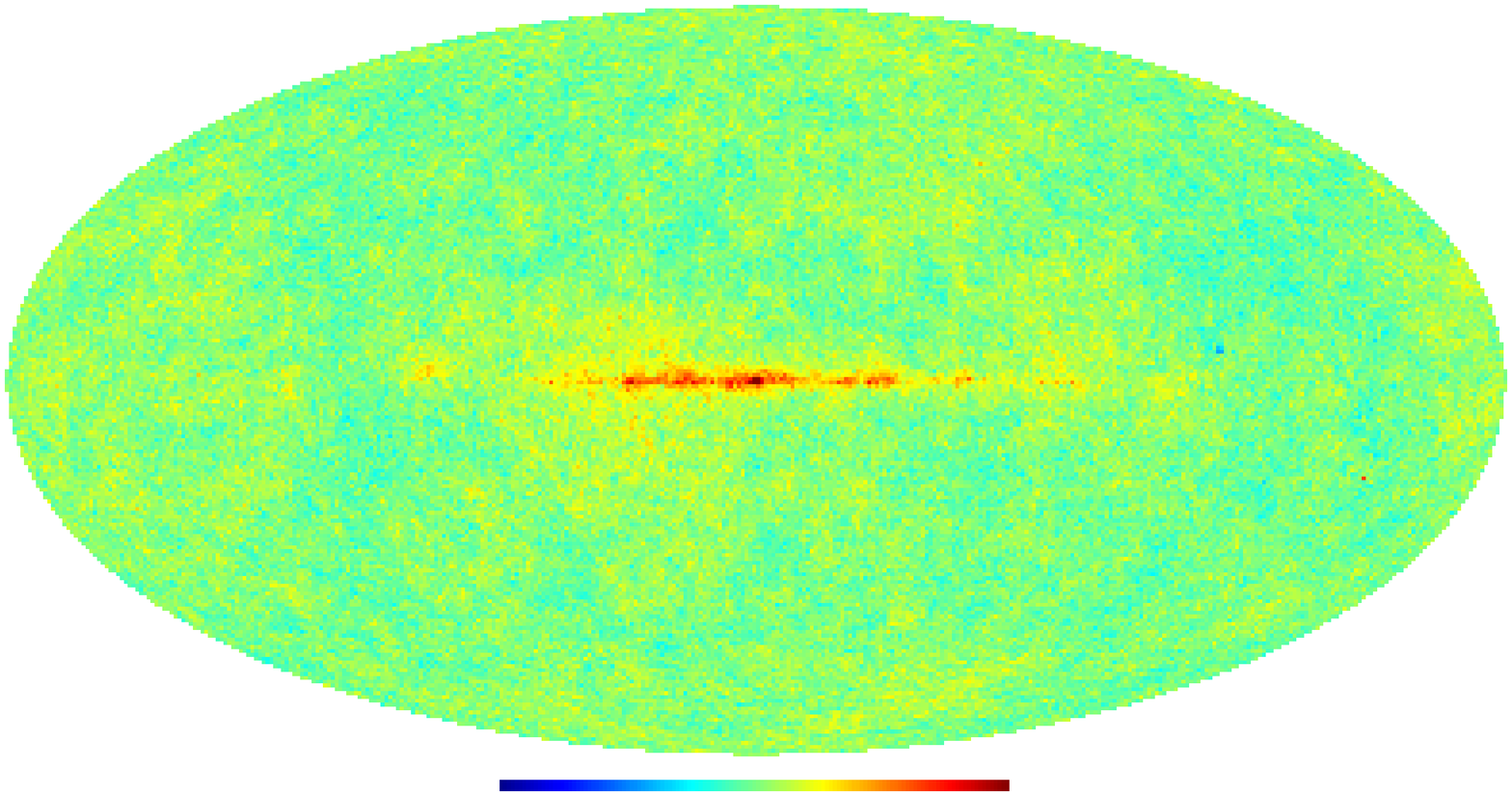}
\includegraphics[draft=false,width=4.7cm,angle=90]{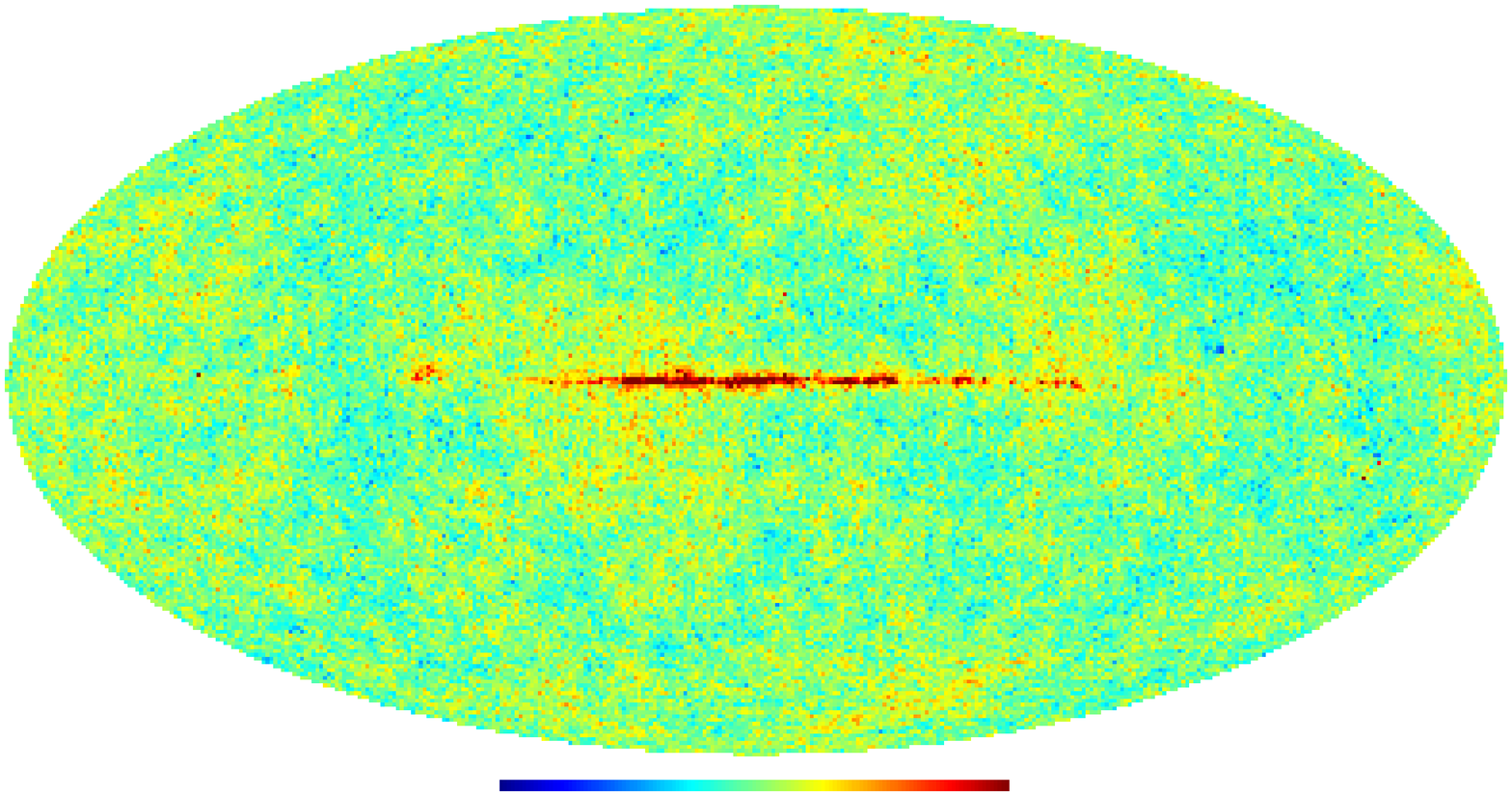}
\includegraphics[draft=false,width=4.7cm,angle=90]{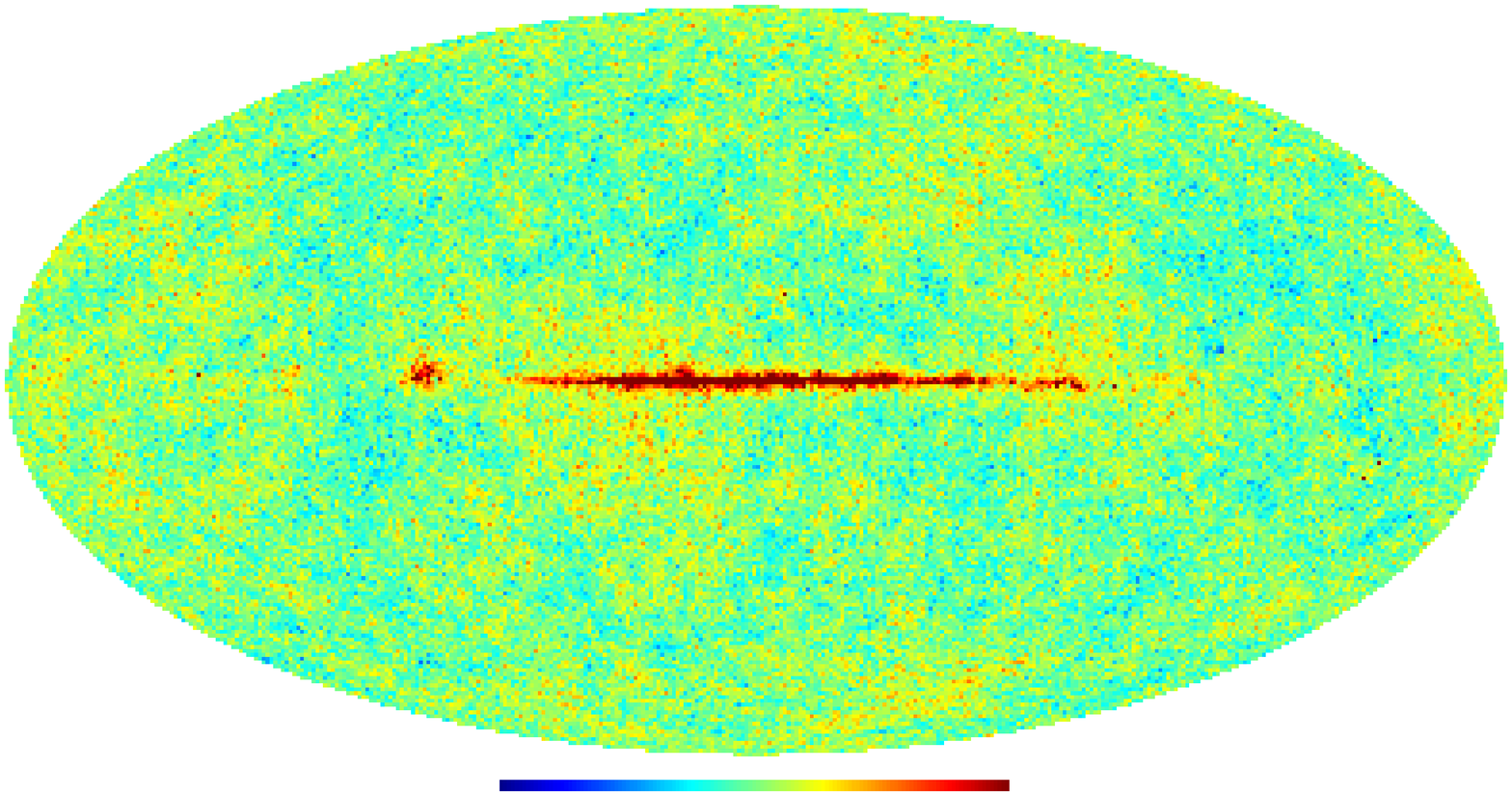}
\includegraphics[draft=false,width=4.7cm,angle=90]{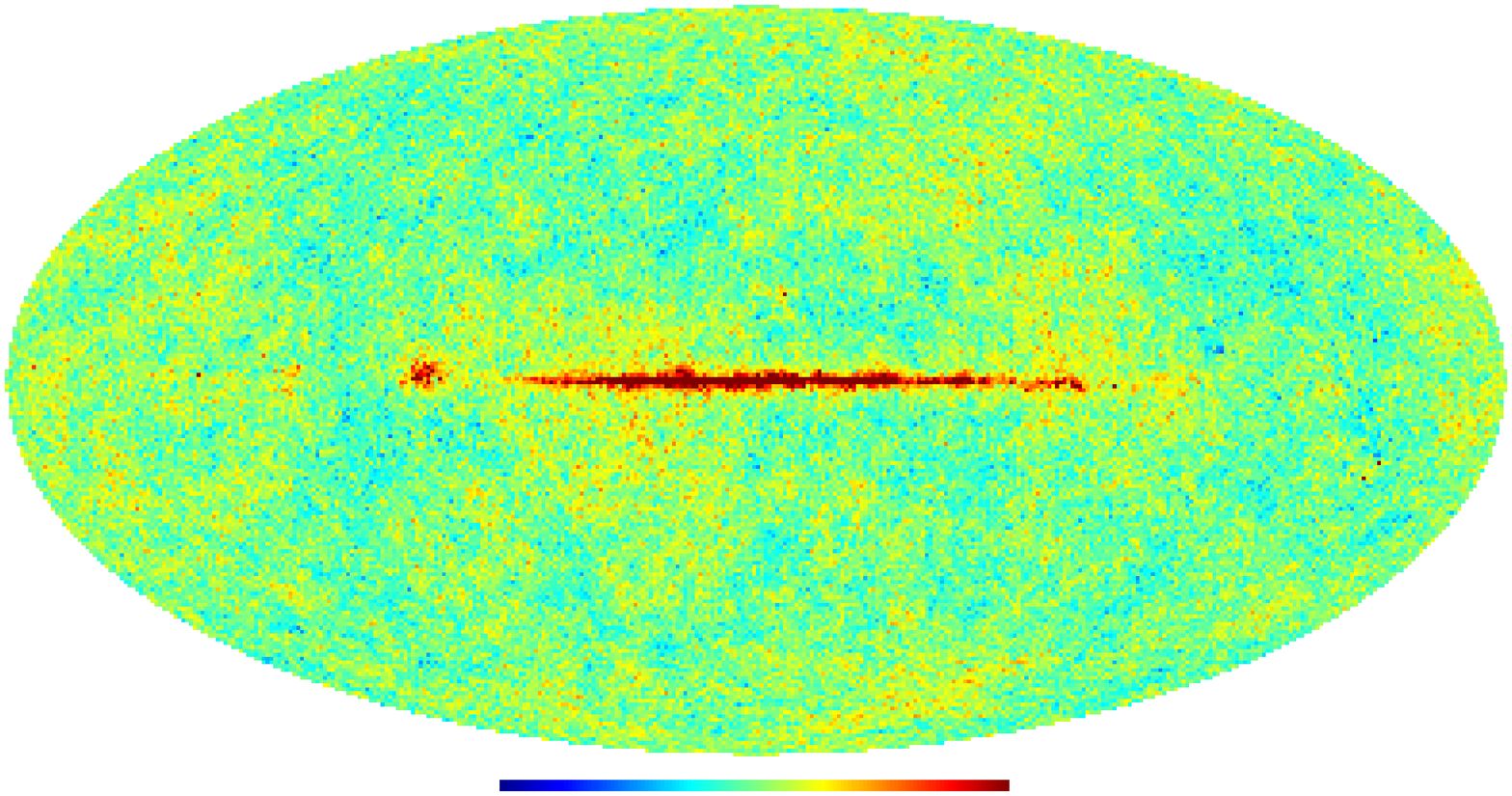}
\includegraphics[draft=false,width=4.7cm,angle=90]{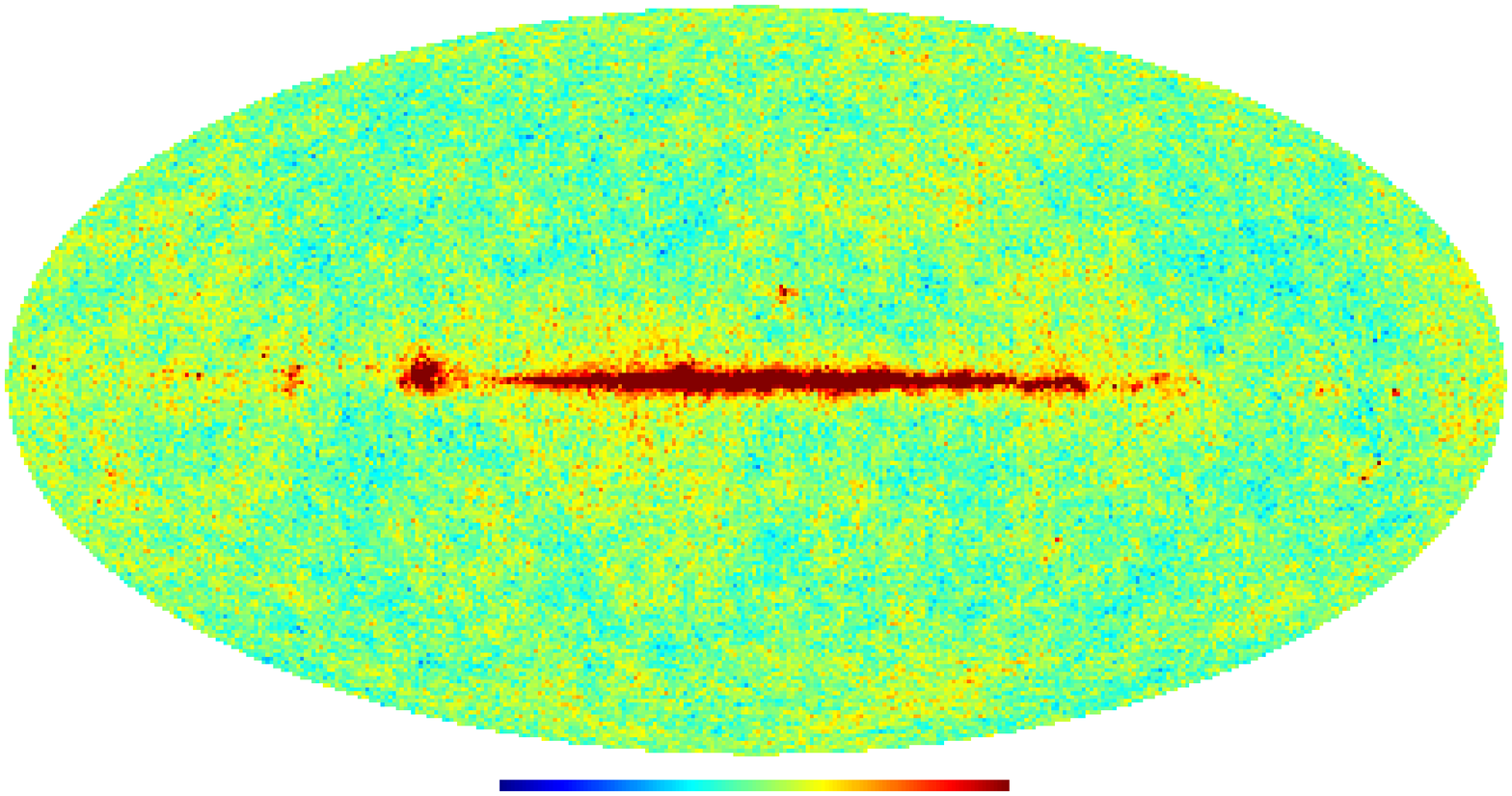}
\includegraphics[draft=false,width=4.7cm,angle=90]{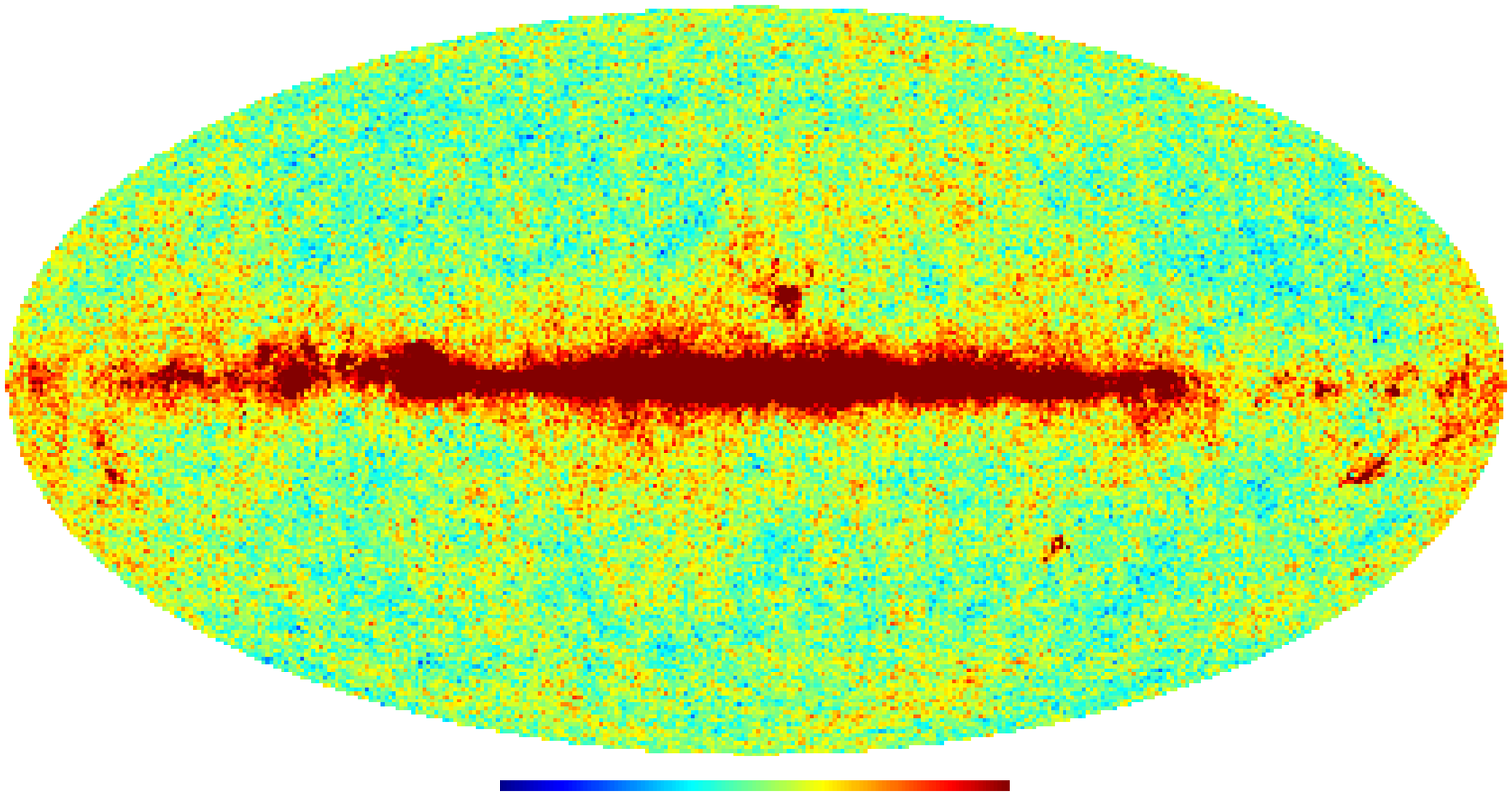}
\includegraphics[draft=false,width=4.7cm,angle=90]{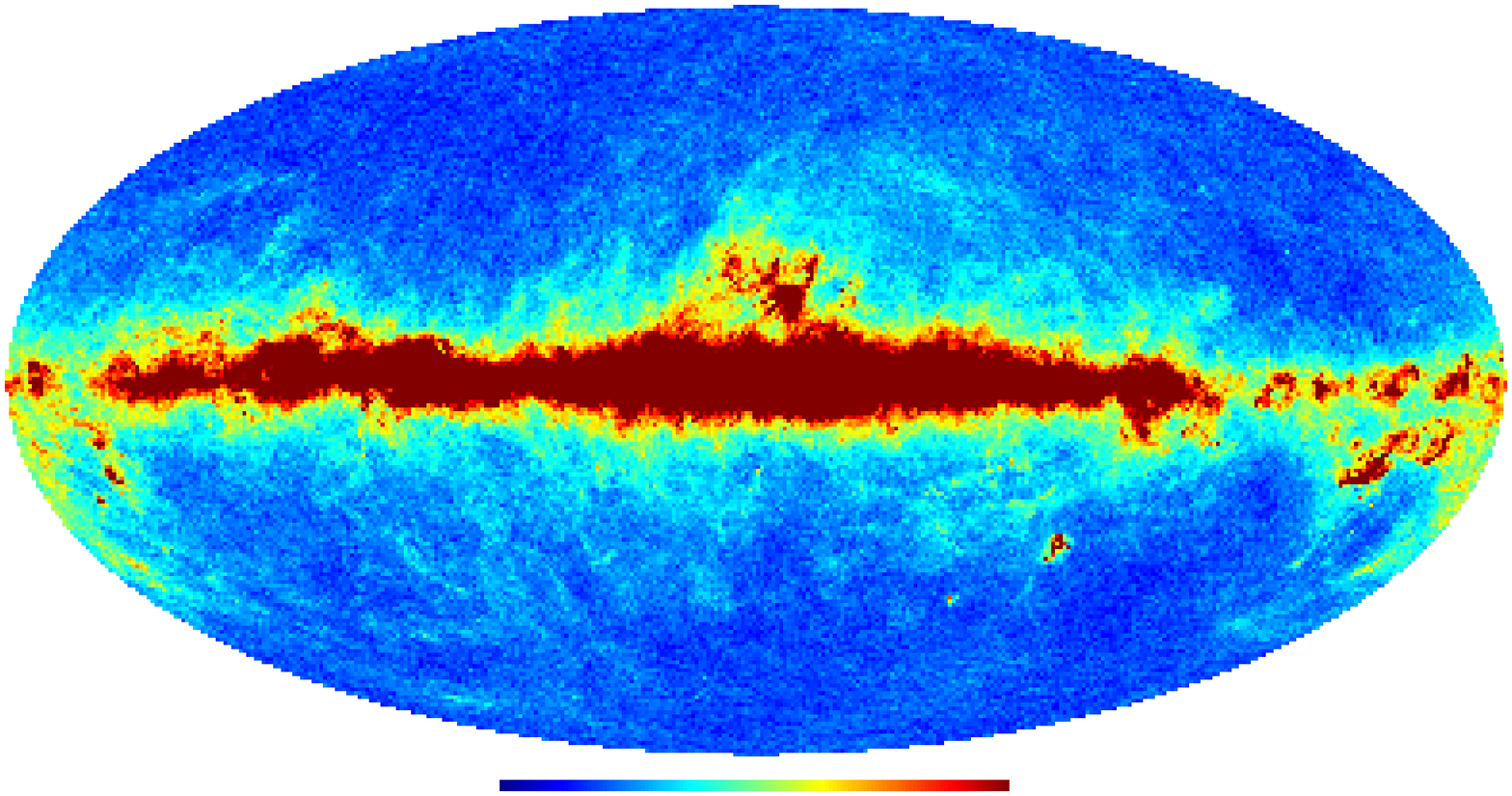}
\includegraphics[draft=false,width=4.7cm,angle=90]{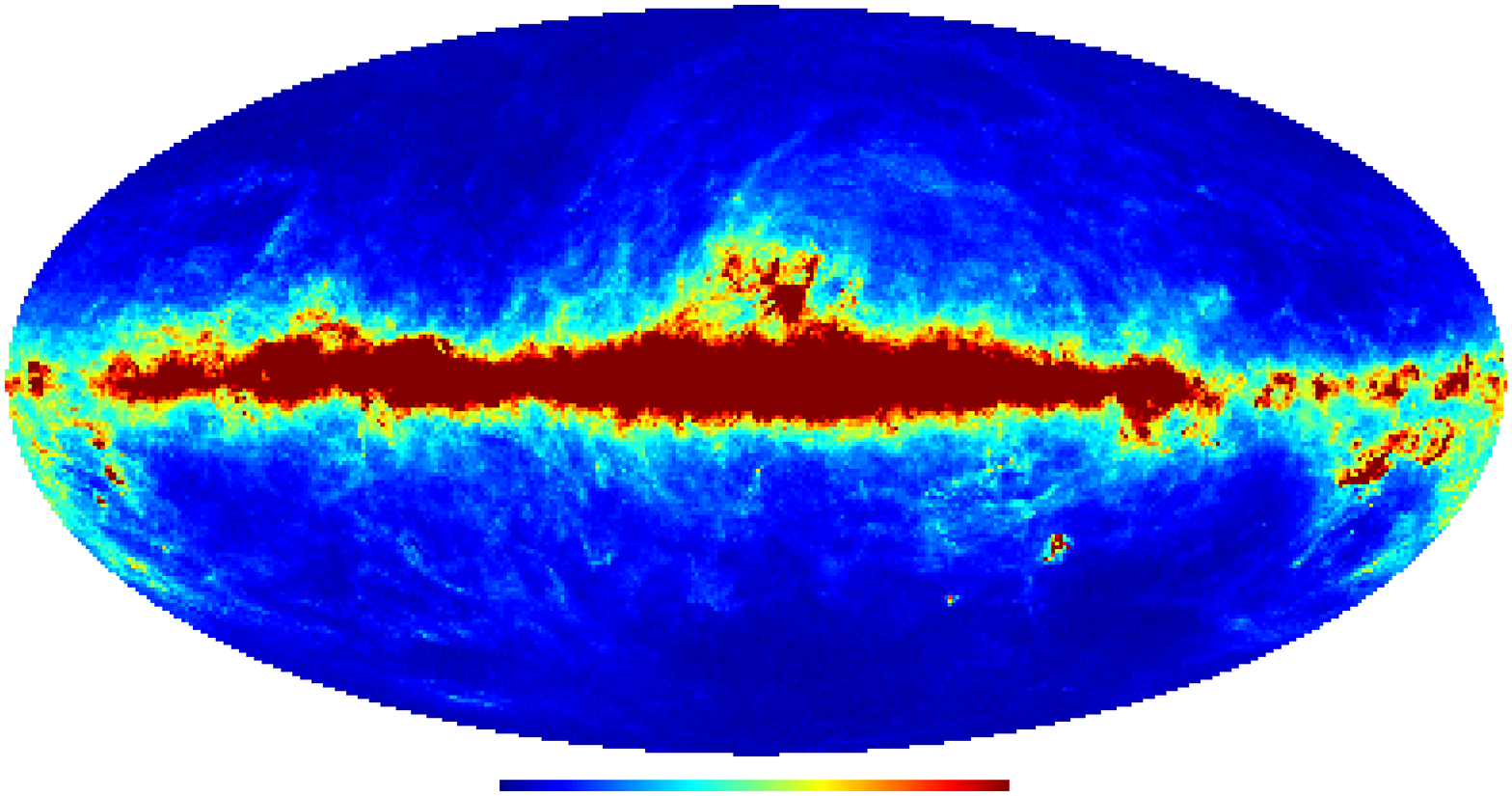}
\includegraphics[draft=false,width=4.7cm,angle=90]{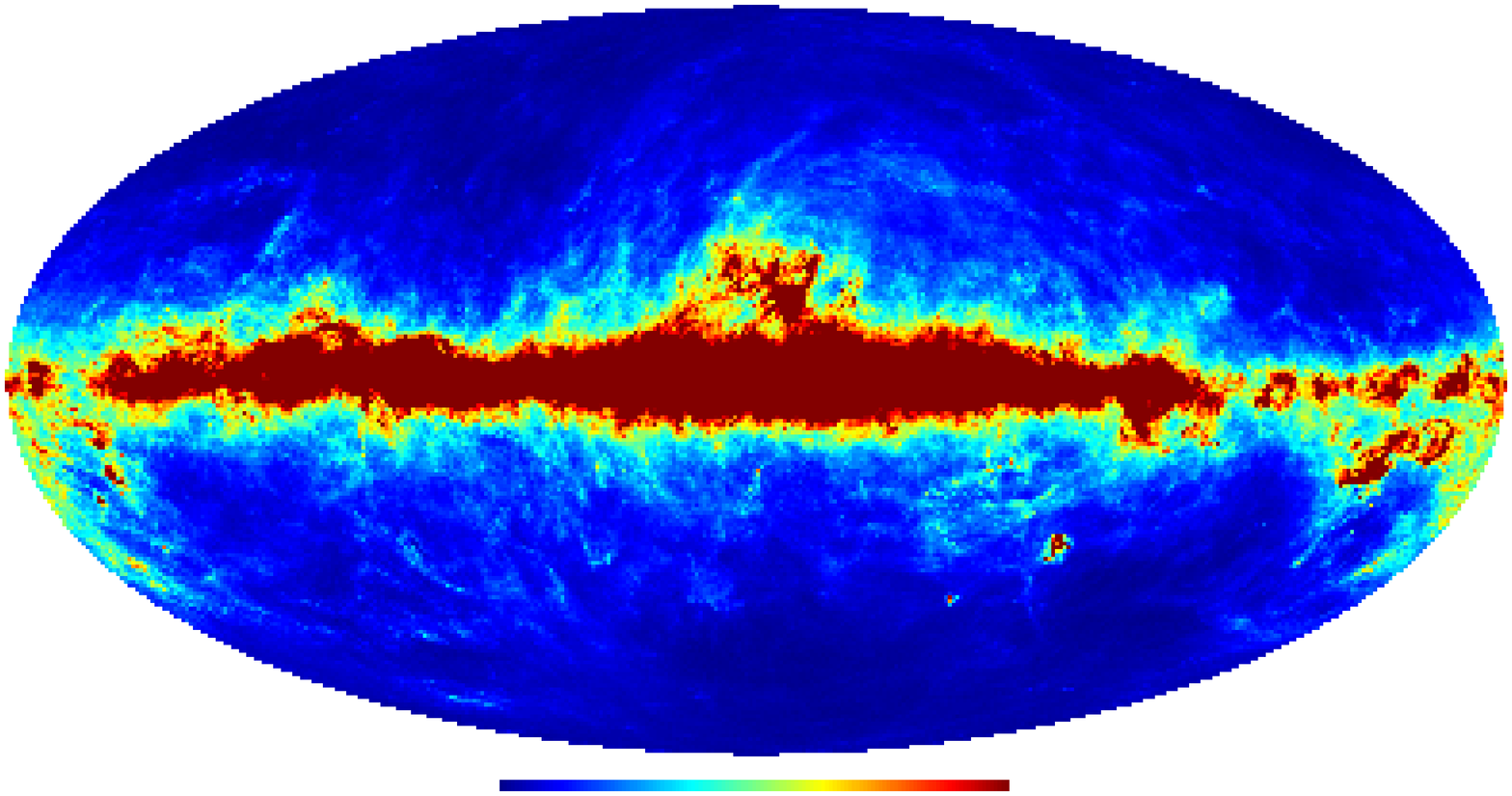}
 \end{center}
\end{minipage}
\label{fig:obsmaps}
\caption{Simulated Planck observation maps at ten frequencies between 30 and 857 GHz}
\end{figure*}

We use full sky Planck simulations provided by the Planck consortium. The maps
are generated at the ten frequencies of Planck instruments (30, 44, 70, 100 GHZ for the low
frequency instrument and 100, 143, 217, 353, 545, 857 GHz for the high frequency
instrument). Five components and white noise at the nominal level of Planck
instruments are mixed according to equation \ref{eq:obsmodel} (the beam sizes are by
increasing order in frequency~: 33, 23, 14, 10, 10.6, 7.4, 4.9, 4.5, 4.5, 4.5
arcminutes). The components are the CMB, the kinetic and thermal SZ effect
from galaxy clusters, and the galactic dust and synchrotron. They are obtained
as follows:\\ The CMB component is randomly generated using a
power spectrum $C_l$ predicted by CMBFast \cite{CMBFAST} with standard cosmological
parameters. The galactic components are generated using observation templates
at different frequencies, the galactic dust is modeled using the DIRBE-IRAS
100$\mu {\rm m}$ maps and the galactic synchrotron is a destriped version of the
408 MHz Haslam survey with additional small scale structures (see \cite{allskyMEM}). The Thermal and
the Kinetic templates are simulated with a gas dynamics code \cite{szsimu}. Note that the
kinetic SZ effect (always sub-dominant) and the CMB component have similar emission
laws. The simulations are performed up to the resolution of 3.5 arcminutes.
Figure \ref{fig:obsmaps} shows the simulated Planck observation maps at all the
frequencies.

\subsection{Estimated parameters}

We have applied our method on the above synthetic data. It
is necessary to fix the number of components assumed to be present in the
data. We choose to characterize 4 components. One of the components in the
simulations, the kinetic SZ, is negligible at all frequencies. Moreover, it
can not be separated from the CMB directly with our approach as the consequence of their proportional
emission law (CMB and kinetic SZ form one component).\\ Spherical harmonics are
computed up to multipole $l$=3000
(corresponding to 3.5 arcminutes), and we choose bins of width $\Delta l = 10$.
We investigate two different approaches~:\\
$\bullet$ First we adopt a quasi-blind approach. We estimate all the elements
of the mixing matrix except three elements, corresponding to
three of the four elements at 857GHz, which are fixed to zero (we expect the presence of only one component at this frequency). We will discuss the
reason for this choice in subsection \ref{subsec:discuss}. All the other parameters are
estimated, including component power spectra and the noise variances. The total
number of parameters is $10 \times 4 -3 +4 \times 300+10 = 1247$, compared to $300 \times 10 \times (10 + 1)/2 = 16500$ data
elements\\
$\bullet$ In a second approach, we assume that we know the electromagnetic
spectra of the components perfectly, so we fix all the elements of the mixing matrix to
their true values. We estimate the component spatial power spectra and the noise
variances.

\subsection{Results}

\begin{figure}[htb]
  \centering
\centerline{\epsfig{figure=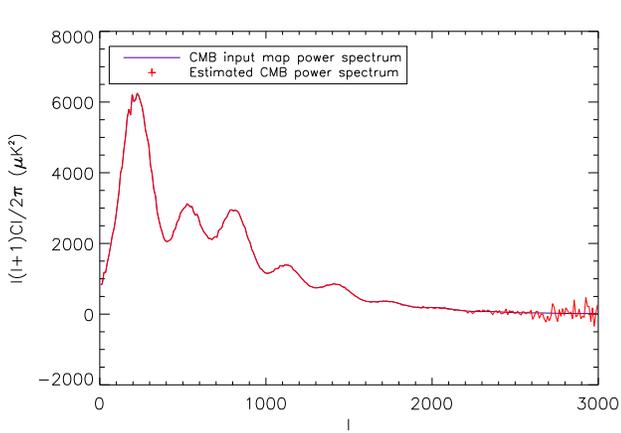,width=8.4cm}}
\caption{Estimated CMB power spectrum in the blind approach}
\label{fig:res0}
\end{figure}

\begin{figure}[htb]
  \centering
\centerline{\epsfig{figure=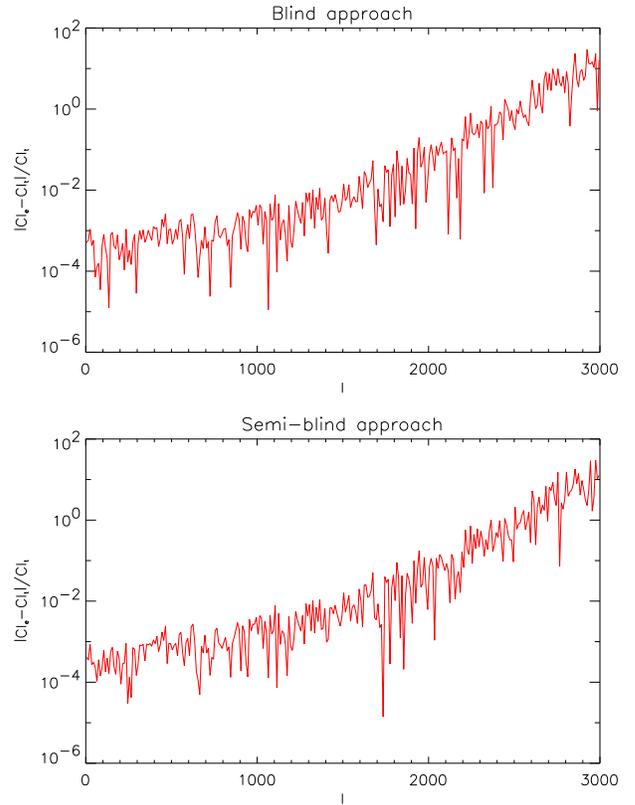,width=8.4cm}}
\caption{Relative errors made on the CMB power spectrum estimation in the (quasi) blind approach (top) and in the semi-blind approach (bottom)}
\label{fig:res1}
\end{figure}

In the blind case, all the parameters of the mixing matrix
corresponding to the CMB and dust components are estimated with a very good
accuracy. 
Table \ref{tab:Arecov} gives the ratio between the recovered and the true
mixing elements of the CMB.
The mixing matrix elements corresponding to the thermal SZ are estimated with
a good precision. The galactic synchrotron emission law is very well constrained at lower
frequencies. Therefore, we show that strong constraints can be put with our method on the
component emission law, in particular for the galactic components
at frequencies far from their maximum emission. \\
\begin{table*}
 \begin{minipage}{2\columnwidth}
   \begin{center}
     \begin{tabular}{@{}lccccccccc}
       channel & 30 & 44 & 70 & 100(LFI) & 100(HFI) & 143  & 217  & 353 & 545\\
       \hline
        CMB & 0.999984 & 1.000254 & 0.999780 & 1.000081 & 1 & 0.999993 & 0.999836 & 0.998972 & 0.990155\\
       \hline
     \end{tabular}
   \end{center}
 \end{minipage}
 \caption{Ratio between the recovered and the true mixing elements of the CMB. The recovered parameters are rescaled to the detector at 100GHz of HFI (which is the best channel for the CMB).}
 \label{tab:Arecov}
\end{table*}
We now concentrate on the spatial power spectrum of the CMB component. Figure \ref{fig:res0} shows the estimated CMB power spectrum in the blind approach. Figure
\ref{fig:res1} shows the relative errors made on the CMB power spectrum estimation given by ($|\tilde C(q) - C(q)|/C(q)$) for the two
approaches.
The remarkable point is that the estimation errors in both approaches are 
equivalent. Therefore, it seems that the emission intensity of the components at the
different observation frequencies can be estimated in Planck data without loss
of accuracy in the linear CMB power spectrum estimation.\\ In both cases, the method
allows to estimate accurately the CMB power spectrum up to $l \simeq 2000$. At smaller scales the dispersion begins to be significant. This
result is not surprising since the noise and the beam smoothing effect
are important at these scales for all detectors. Also, the estimated power
spectrum does not seem to be contaminated by the kinetic SZ.

\subsection{Discussion}
\label{subsec:discuss}


{\bf Semi-blind approach?}

As we have seen before, in the case where we estimate the mixing matrix
as well as the power spectra of the components, we have fixed to zero the
contribution of three components at 857GHz, the only component we assume to be
present at this frequency is the galactic dust. This
hypothesis, being very realistic, appears to be necessary because we expect that
the galactic dust and the galactic synchrotron have quasi proportional power
spectra. Without any prior, the method can not separate these two components,
but the previous simple procedure of fixing some parameters allows to break
the degeneracies and to estimate accurately all the components.

{\bf Partial coverage}

The galactic components are very inhomogeneous on the sky. In the region of
the galactic plane, their intensities are several hundred times stronger than in
the outer regions. Since we model the components as homogeneous, the method, when applied on those data, is sub-optimal. However, more accurate CMB power
spectrum estimation may be obtained by making a galactic cut, involving a
partially covered sky processing. The coefficients obtained from the spherical
harmonic decomposition of partially covered sky are correlated. Our
spectral matching method can be used on such data and a good choice would be to take bin
size $\Delta l$ larger than the correlation length in the power spectrum measurement.
The method, in particular, is applicable to data such that from the Archeops balloon-borne
experiment in which the sky coverage is about $30\%$ of the whole sky.

\section{Conclusion}
\label{sec:conclusion}

We have adapted our blind MDMC spectral matching method for the processing of all sky CMB
maps. The observations are modeled as a noisy linear mixture of beam-convolved components.
By maximizing the likelihood of the system, we estimate the power spectra of 
the components, their contribution levels at each frequency as well as the noise
levels. The method has been applied on full sky Planck simulations
containing five components. The spherical harmonics basis was used.\\ The power spectrum of the CMB is accurately estimated up to $l \simeq 2000$ in bins of
size $\Delta l = 10$.
The comparison between the results obtained in blind and semi-blind approaches
shows that the mixing elements can be estimated without loss of accuracy in
the CMB power spectrum estimation. 

\section{Acknowledgments}
We thank the Planck collaboration and in particular V. Stolyarov and R. Kneissl
for the full sky simulated maps.
Guillaume Patanchon would like to thanks M.A.J. Ashdown for
useful discussions while he was visiting Cavendish Laboratory.
The HEALPix package \cite{healpix} (see http://www.eso.org/ science/healpix/) was used for the spherical harmonics decomposition of the input maps.


\bibliographystyle{IEEEbib}
\bibliography{PSIP2003for_astroph}







\end{document}